\documentclass[a4paper]{jpconf}
\usepackage{graphicx}
\begin{document}
\title{The quantum formulation derived from assumptions of epistemic processes}

\author{Inge S. Helland}

\address{Professor emeritus, Department of Mathemathics, University of Oslo}

\ead{ingeh@math.uio.no}

\begin{abstract}
Motivated by Quantum Bayesianism I give background for a general epistemic approach to quantum mechanics, where complementarity and symmetry are the only essential features. A general definition of a symmetric epistemic setting is introduced, and for this setting the basic Hilbert space formalism is arrived at under certain technical assumptions. Other aspects of ordinary quantum mechanics will be developed from the same basis elsewhere.
\end{abstract}

\section{Introduction}

The ordinary textbook formulation of quantum mechanics is very abstract. Its starting point: 'The state of a physical system is a normalized vector in a separable Hilbert space' has lead to an extremely rich theory, a theory which has not been refuted by any experiment and whose predictions range over an extremely wide varity of situations. Nevertheless, it is still unclear how this state concept should be interpreted.

Many conferences on quantum foundation have been arranged in recent years, but this has only implied that the number of new interpretations have increased, and no one of the old have died out. In two of these conferences, a poll among the participants was carried out [1, 2]. The result was an astonishing disagreement on several simple and fundamental questions. One of these questions was whether quantum theory should be interpreted as an objective theory of the world (the ontological interpretation) or if it only expresses our knowledge of the world (the epistemic interpretation). According to Webster's Unabridged Dictionary, the adjective 'epistemic' means 'of or pertaining to knowledge, or the conditions for acquiring it'.

In Spekkens [3] a toy model was developed, a toy model which can be taken as a strong argument in favour of an epistemic interpretation of quantum mechanics. Partly based on this and partly on other basic premises, Fuchs [4, 5, 6] and others have argued for various versions of Quantum Bayesianism, a radical interpretation where the subjective observer plays an important part. See also the philosophical discussion by Timpson [7] and the popular account in [8].

The present paper takes as a point of departure that in some sense or other the epistemic interpretation should be important for the fundametal issues of the quantum world. The next question then arises: Can one find a new and more intuitive \textit{foundation} of quantum theory, a foundation related to the epistemic interpretation? It is my view that such a foundation must have some relation to statistical inference theory, another scientific fundament, which gives tools for   wide variety of empirical investigations, and which in its very essence is epistemic. It should also be a kind of decision theory, related to decisions taken in everyday life. 

Every epistemic question together with an answer to this question imply at least two decisions: First a decision to focus in the selection of the question to ask. Then data are collected, and one ends up with a decision on which answer to the focused question one should give from these data. This last decision is covered by traditional decision theory, but one needs to pay more attention to the process of focusing.

A very tentative answer to the question on foundation given above, is then yes, but there is still more work to be done before such a foundation can be completely established. In [9] I will aim to carry out the arguments in more detail.

\section{Background and definitions}

The Quantum Bayesianism is founded on an observer's belief, quantified by a Bayesian probability. I want to relate my state concept to the notion of \emph{certain} belief, which I call knowledge. The knowledge will be associated with an agent or with a group of communicating agents, and his/her/their knowledge will be knowledge about what I will call an e-variable.

An epistemic process is any process under which an agent or a group of communicating agents obtain knowledge about a physical system. In general there are many ways by which one can obtain knowledge about the world or about aspects of the world. In a given situation the observer has some background, in terms of his history, in terms of his physical environment, and in terms of the concepts that he is able to use in analyzing the situation. All this may limit his ability to obtain knowledge.

A \emph{conceptual variable} is any variable related to the physical system, defined by an agent or by a group of communicating agents. The variable may be a scalar, a vector or belong to a larger space.

An \emph{e-variable} or epistemic conceptual variable $\theta$ is a conceptual variable associated with an epistemic process: Before the process the agent (or agents) has (have) no knowledge about $\theta$; after the process he/they has/have some knowledge, in the simplest case full knowledge: $\theta=u_k$. Here $u_k$ is one of the possible values that $\theta$ can take. In this paper it is mostly assumed for simplicity that $\theta$ is discrete, which it will be in the elementary quantum setting below. For a continuous variable $\theta$, knowledge on the e-variable will be taken to mean a statement to the effect that $\theta$ belongs to some given set, an interval if $\theta$ is a scalar.

The e-variable concept is a generalization of the parameter concept as used in statistical inference, introduced by Fisher [10], and today incorporated in nearly all applications of statistics. In statistics, a parameter $\theta$ is usually an index in the statistical model for the observations, and the purpose of an empirical investigation is to obtain statements about $\theta$, in terms of point estimation, confidence interval estimation or conclusion from the testing of hypotheses. The parameter is often associated with a hypothetical infinite population. My e-variable will also be allowed to be associated with a finite physical system, a particle or a set of particles. But, in the same way as with a parameter, the purpose of any empirical investigation will be to try to conclude with some statement, a statement expressed in terms of an e-variable $\theta$.

It is important that not all conceptual variables are e-variables. A conceptual variable $\phi$ is called \emph{inaccessible} if there is no epistemic process by which one can get accurate knowledge about it. An example from the area of quantum mechanics is $\phi=(\xi,\pi)$, where $\xi$ is the position of a particle, and $\pi$ is the momentum. Also macroscopic examples abound, for instance connected to counterfactual situations [9]. Another physical example is $\phi=(\lambda^x ,\lambda^y ,\lambda^z )$, where $\lambda^a$ is the component of an angular momentum or a spin for a particle or a system of particles, the component in direction $a$. Here each $\lambda^a$ is an accessible conceptual variable, an e-variable, but the vector $\phi$ is inaccessible.

In such cases, where a vector of e-variables is inaccessible, it is equivalent to say that the components are complementary. Since introduced by Bohr, the concept of complementarity has played a fundamental and important role in quantum mechanics.

It is essential to stress that the inaccessible conceptual variables above are not hidden variables. Variables like $\phi$ are just mathematical variables, but variables upon which group actions may be defined. $\phi=(\xi,\pi)$ may be subject to Galilean transformations, time translations or changes of units, while $\phi=(\lambda^x ,\lambda^y ,\lambda^z )$ may be subject to rotations. This will of course also induce transformation of the components, the e-variables. Important transformations in the group of rotations of $\phi$ are: 1) Those leading to the change in the values of $\lambda^x$ (or of any other fixed component); 2) Those leading to an exchange of $\lambda^x$ and $\lambda^y$ (or any other pair of components).

\section{The maximal symmetrical epistemic setting}

The purpose of this paper is to indicate how one can derive the essence of the quantum formalism from reasonable assumptions in a setting which generalizes the angular momentum setting described above. 

Let in general $\phi$ be an inaccessible conceptual variable taking values in some measurable space $(\Phi, \mathcal{F})$, and let $\lambda^a =\lambda^a (\phi)$ be accessible 
functions for $a$ belonging to some index set $\mathcal{A}$. I will repeat that a conceptual variable is accessible if it in the given context can be estimated with arbitrary 
accuracy by \emph{some} experiment. Technically I will without further mention assume that all functions defined on $\Phi$ are 
measurable. To begin with, I will assume that the functions $\lambda^a$ are maximal, and also that they can be transformed into each other by an automorphism in $\Phi$.
\bigskip

\textbf{Assumption 1} \textit{ a) Consider the partial ordering defined by $\alpha <\beta$ iff $\alpha =f(\beta)$ for some function $f$. Under this partial 
ordering each $\lambda^a (\phi)$ is maximally accessible.}

\textit{b) For $a\ne b$ there is an invertible transformation $g_{ab}$ such that $\lambda^b (\phi)=\lambda^a (g_{ab}\phi)$.}
\bigskip

Note that the partial ordering in a) is consistent with accessibility: If $\beta$ is accessible and $\alpha=f(\beta)$, then $\alpha$ is accessible. Also, $\phi$ is an 
upper bound under this partial ordering. The existence of maximal accessible conceptual variables follows then from Zorn's lemma.

For the angular momentum case, the transformation $g_{ab}$ means exchange of the component $\lambda^a$ and the component $\lambda^b$. In this  process, one can imagine that the vector $\phi$ is decomposed as $\phi=(\lambda^a,\lambda^b,\lambda^c)$, where the axis $c$ is orthogonal to the $(a,b)$-plane, and $g_{ab}$ then changes the sign of $\lambda^c$, moving along a great circle.

Below, I will often single out a particular index $0\in\mathcal{A}$. Then a), given b), can be formally weakened to the assumption that $\lambda^0 (\phi)$ is 
maximally accessible, and b) can be weakened to the existence for all $a$ of an invertible transformation $g_{0a}$ such that 
$\lambda^a (\phi)=\lambda^0 (g_{0a}(\phi))$.  Take $g_{ab}=g_{0a}^{-1}g_{0b}$. 

Even though $\phi$ is inaccessible, it is possible to operate on $\phi$ with functions, in particular group actions. The group of automorphisms on $\Phi$, all 
transformations of elements $\phi$, always exists from a mathematical point of view, and one can imagine many subgroups of this group. Some of these 
will now be defined.
\bigskip

\textbf{Definition 1} \textit{For each $a$, let $\tilde{G}^a$ be the group of automorphisms on $\Lambda^a$, the space upon which $\lambda^a$ varies. 
For $\tilde{g}^a \in \tilde{G}^a$ let $g^a$ be any transformation on $\Phi$ for which $\tilde{g}^a \lambda^a (\phi)=\lambda^a (g^a \phi)$.}
\bigskip

It is easily verified that 1) For fixed $\tilde{g}^a$ the transformations $g^a$ form a set which is invariant under decompositions and inversion. 2) For fixed $a$ the transformations $g^a$ form a group $G^a$.

Thus the group $G^a$ is the group transforming values of $\lambda^a$ into the same or other values of this e-variable, and the corresponding group 
$\tilde{G}^a$ is the group of all transformations of these values. 

Obvious consequences of Definition 1 are that $\tilde{G}^a$ is transitive over $\Lambda^a$ and that $G^a$ is transitive over $\Phi$.

Now single out a fixed index $0\in \mathcal{A}$.
\bigskip

\textbf{Definition 2} \textit{Let $G$ be the group of transformations generated by $G^0$ and the transformations $g_{0a}$, $a\in \mathcal{A}$.}
\bigskip

It is easily verified that $G^a = g_{0a}^{-1} G^0 g_{0a}.$ Together with $g_{ab}=g_{0a}^{-1}g_{0b}$ this implies that $G$ also is the group generated by 
$G^a$, $a\in \mathcal{A}$ and $g_{ab}$, $a,b \in \mathcal{A}$.

Now I want to introduce the further
\bigskip

\textbf{Assumption 2} \textit{a) The group $G$ is a locally compact topological group, and satisfies weak assumptions such that an invariant 
measure $\rho$ on $\Phi$ exists.}

\textit{b) The group generated by products of elements of $G^a ,G^b ,...$; $a,b,...\in\mathcal{A}$ is equal to $G$.}
\bigskip

Assumption 2a) is a technical one, needed in the next section. Note that $G$ is defined in terms of transformations upon $\Phi$, so that the topology must be introduced 
in terms of these transformations. Technically this can for instance be achieved by assuming $\Phi$ to be a metric space with metric $d$, and letting $g_n\rightarrow g$ if 
$\mathrm{sup}_\phi d(g_n \phi, g\phi)\rightarrow 0$. Concerning Assumption 2b), it follows from 
$g^a g^b ...=g_{0a}^{-1}g^{0}g_{0a}g_{0b}^{-1}g^{0'}g_{0b},...$, where $g^a \in G^a , g^b \in G^b ,...$ and $g^{0}, g^{0'},...\in G^0$, that the group of 
products is contained in $G$. That it is equal to $G$, is an assumption on the richness of the set $\mathcal{A}$ or the richness of $G^0$.

Here is a proof that assuption 2b) holds in the angular momentum/ spin case: Let $c$ be any direction perpendicular to the midline between $a$ and $b$. Let $g^c$ be the group element changing the sign of $\lambda^c$, the $c$-component of $\phi$, moving along a great cicle. Then $g_{ab}=g^c$.

The setting described here, where Assumption 1 and Assumption 2 are satisfied, includes many quantum mechanical situations including angular momenta and spins. 
I will call it \emph{the maximal symmetrical epistemic setting}. 

An important special case is when each $\Lambda^a$ is discrete. Then $\tilde{G}^a$ is the group of permutations of elements of $\Lambda^a$.

\section{Quantum reconstruction}

Fix $0\in\mathcal{A}$, and let $H$ be the Hilbert space
\[H=\{f\in L^2(\Phi,\rho):\ f(\phi)=\tilde{f}(\lambda^0(\phi))\ \mathrm{for\ some\ }\tilde{f}\}.\]
I will assume that the $\lambda^a$'s are discrete. Then $H$ is separable. If the $\lambda^a$'s take $d$ different values, $H$ is $d$-dimensional. Since all separable Hilbert spaces are isomorphic, it is enough to arrive at the quantum formulation on this $H$.
\bigskip

\textbf{Lemma 1} \textit{The values $u_k^a$ of $\lambda^a$ can always be arranged such that $u_k^a=u_k$ is the same for each $a$ ($k=1, 2, ...$).}
\bigskip

\underline{Proof.} By Assumption 1
\[\{\phi:\lambda^b=u_k^b\}=\{\phi:\lambda^a(g_{ab}\phi)=u_k^b\}=g_{ba}(\{\phi:\lambda^a(\phi)=u_k^b\}).\]
The sets in brackets on the lefthand side here are disjoint with union $\Phi$. But then the sets in brackets on the righthand side are disjoint with union $g_{ab}(\Phi)=\Phi$, and this implies that $\{u_k^b\}$ gives all possible values of $\lambda^a$.
\smallskip

Now I am able to formulate the main result of this paper:
\bigskip

\textbf{Theorem 1} \textit{a) For every $a$, $u_k$ and associated with every indicator function $I(\lambda^a (\phi)=u_k)$ there is a vector $|a;k\rangle\in H$, which is unique in the sense that $|a;k\rangle\ne|b;j\rangle$ for all $a, b, j, k$ except in the trivial case $a=b, j=k$. This inequality is interpreted to mean that there is no phase factor $e^{i\gamma}$ such that $|a;k\rangle=e^{i\gamma}|b;j\rangle$.}

\textit{b) For each $a$ the vectors $|a;k\rangle$ form an orthonormal basis for $H$.}
\bigskip

This gives us the possibility to interprete the vectors $|a;k\rangle$, corresponding to the indicators $I(\lambda^a (\phi)=u_k)$, as follows:
\smallskip

\textit{1) The question 'What is the value of $\lambda^a$?' has been focused on.} 

\textit{2) Through an epistemic process we have obtained the answer '$\lambda^a=u_k$'.}
\smallskip

When a ket vector is defined, a corresponding bra vector can be defined. The operator corresponding to $\lambda^a$ can be defined as
\[A^a =\sum_k u_k |a;k\rangle\langle a;k|.\]
In the maximal setting this has non-degenerate eigenvalues.

\section{Proof under an extra assumption}

Let $U$ be the left regular representation of $G$ on $L^2(\Phi,\rho)$: $U(g)f(\phi)=f(g^{-1}\phi)$. It is well known that this is a unitary representation. We will seek a corresponding representation of $G$ on the smaller space $H$.

In the following, recall that upper indices as in $g^a$ indicate variables related to a particular $\lambda^a$, here a group element of $G^a$. Also recall that $0$ is a fixed index in $\mathcal{A}$. Lower indices as in $g_{ab}$ has to do with the relation between two different $\lambda^a$ and $\lambda^b$.
\bigskip

\textbf{Proposition 1} \textit{a) A (multivalued) representation $V$ of $G$ on the Hilbert space $H$ can always be found.}

\textit{b) There is an extended group $G'$ such that $V$ is a univalued representation of $G'$ on $H$.}

\textit{c) There is a homomorphism $G'\rightarrow G^0$ such that $V(g')=U(g^0)$. If $g'\ne e'$ in $G'$, then $g^0\ne e$ in $G^0$.}
\bigskip

\underline{Proof.} a) For each $a$ and for $g^a\in G^a$ define $V(g^a)=U(g_{0a})U(g^a)U(g_{a0})$. Then $V(g^a)$ is an operator on $H$, since it is equal to $U(g_{0a}g^a g_{a0})$, and $g_{0a}g^a g_{a0}\in G^0 =g_{0a}G^a g_{a0}$. For a product $g^a g^b g^c$ with $g^a\in G^a$, $g^b \in G^b$ and $g^c \in G^c$ we define $V(g^a g^b g^c)=V(g^a )V(g^b)V(g^c)$, and similarly for all elements of $G$ that can be written as a finite product of elements from different subgroups. 

Let now $g$ and $h$ be any two elements in $G$ such that $g$ can be written as a product of elements from $G^a, G^b$ and $G^c$, and similarly $h$ (the proof is 
similar for other cases.) It follows that $V(gh)=V(g)V(h)$ on these elements, since the last factor of $g$ and the first factor of $h$ either must belong to 
the same subgroup or to different subgroups; in both cases the product can be defined by the definition of the previous paragraph. In this way we see that $V$ is 
a representation on the set of finite products, and since these generate $G$ by Assumption 2b) , it is a representation of $G$.

Since different representations of $g$ as a product may give different solutions, we have to include the possibility that $V$ may be multivalued.
\smallskip

b) Assume as in a) that we have a multivalued representation $V$ of $G$. Define a larger group $G'$ as follows: If $g^a g^b g^c =g^d g^e g^f$, say, 
with $g^k \in G^k $ for all $k$, we define $g_1'=g^a g^b g^c $ and $g_2'=g^d g^e g^f$. Let $G'$ be the collection of all such new elements that can be written as a formal product of elements $g^k \in G^k$. 
The product is defined in the natural way, and the inverse by for example $(g^a g^b g^c )^{-1}=(g^c )^{-1}(g^b )^{-1}(g^a)^{-1}$. By Assumption 2b), the group $G'$ 
generated by this construction must be at least as large as $G$. It is clear from the proof of a) that $V$ also is a representation of the larger group $G'$ on $H$, now a one-valued representation.
\smallskip

c)  Consider the case where $g' =g^a g^b g^c$ with $g^k \in G^k$. Then by the proof of a):
\[V(g') = U(g_{0a}) U(g^a)U(g_{a0}) U(g_{0b}) U(g^b)U(g_{b0})U(g_{0c}) U(g^c)U(g_{c0}) \]
\[=U(g_{0a}g^a g_{a0} g_{0b}g^b g_{b0} g_{0c} g^c g_{c0})
=U(g^0),\]
where $g^0 \in G^0$. The group element $g^0$ is unique since the decomposition $g'=g^a g^b g^c$ is unique for $g'\in G'$. The proof is similar for other decompositions. By the construction, the mapping $g'\rightarrow g^0$ is a homomorphism.

Assume now that $g^0 =e$ and $g'\ne e'$. Since $U(g^0)\tilde{f}(\lambda^0(\phi))=\tilde{f}(\lambda^0((g^0)^{-1}(\phi)))$, it follows from $g^0=e$ that $U(g^0)=I$ on $H$. 
But then from what has been just proved, $V(g')=I$, and since $V$ is a univariate representation, it follows that $g'=e'$, contrary to the assumption.
\bigskip

Now choose an orthonormal basis for $H$: $f_1,...,f_d$ where $f_k(\phi)=\tilde{f}_k(\lambda^0(\phi))$, and where the interpretation of $f_k$ is that $\lambda^0=u_k$. Write $|0;k\rangle=f_k(\phi)$.

Introduce the assumption that the representation $V$ really is multivalued, in a sense to be made precise below. Let $g_{0a1}'$ and $g_{0a2}'$ be two different elements of the group $G'$, both corresponding to $g_{0a}$ of $G$. Define $g_a'=(g_{0a1}')^{-1}g_{0a2}'$. Then $g_a'\ne e'$ in $G'$. By the homomorphism of Proposition 1c), let $g_a'\rightarrow g_a^0$. Then $g_a^0\ne e$ in $G^0$. Now define
\[|a;k\rangle = \tilde{f}_k(\lambda^0(g_a^0\phi))=U((g_a^0)^{-1})|0;k\rangle.\]

\textbf{Assumption 3}
a) $g_{0a1}'$ and $g_{0a2}'$ can be chosen so that $\tilde{g}_a^0\lambda\ne\lambda$ for all $\lambda\in\Lambda^0$ when $a\ne 0$.

b) $g_{0a1}', g_{0a2}', g_{0b1}'$ and $g_{0b2}'$ can be chosen so that $\tilde{g}_a^0\lambda\ne \tilde{g}_b^0\lambda$ for all $\lambda\in\Lambda^0$ when $a\ne b$.
\bigskip

\underline{Proof of Theorem 1 under Assumption 3.}
Let $j\ne k$. I will first prove that the basis functions $f_1,...,f_d$ can be chosen so that $|a;k\rangle\ne |0;j\rangle$ for all $a$. Here and below, inequality of state vectors is interpreted to mean that they can not be made equal by introducing a phase factor. To this end, choose $\tilde{f}_j$ and $\tilde{f}_k$ in such a way that there exists an $\lambda_0^j$ such that $\tilde{f}_k(\lambda)\ne\tilde{f}_j(\lambda_0^j)$ for all $\lambda\in \Lambda^0$. A possible choice is the indicator function $\tilde{f}_j(\lambda)=I(\lambda=u_j)$, but there are other choices. Then for any fixed $g$, $\tilde{f}_{kg}$ defined by $\tilde{f}_{kg}(\lambda^0(\phi))=\tilde{f}(\lambda^0(g\phi))$ is different from $\tilde{f}_j$, and $|a;k\rangle\ne |0;j\rangle$ for all $a$.

Next fix $k$ and choose $\tilde{f}_k$ in such a way that there exists a $\lambda_1^k$ such that $\tilde{f}_k(\lambda)\ne\tilde{f}_k(\lambda_1^k)$ when $\lambda\ne\lambda_1^k$ and $\lambda\in \Lambda^0$. A possible choice is again $\tilde{f}_k(\lambda)=I(\lambda=u_k)$. For $|a;k\rangle\ne|0;k\rangle$ it is sufficient that $\tilde{f}_k(\lambda^0(g_{a}^0\phi))\ne\tilde{f}_k(\lambda^0(\phi))$ for at least one $\phi$, and this will hold if $\lambda^0(g_a^0\phi)\ne\lambda^0(\phi)$ for all $\phi$. This is equivalent to Assumption 3a).

The proof that $|a;k\rangle\ne |b;j\rangle$ (except in the trivial case $a=b, k=j$) holds under assumption 3b), is similar.

The vectors $|0;k\rangle$ are chosen to be an orthonormal basis for $H$. Since $|a;k\rangle=U|0;k\rangle$ for some unitary $U$, it follows that the vectors $|a;k\rangle$ form an orthonormal basis.
\bigskip

Assumption 3 is not satisfied for the spin/ angular momentum case. Nevertheless it is shown in [16] that Theorem 1 holds also for this case.

\section{The general symmetrical epistemic setting}

Go back to the definition of the maximal symmetrical epistemic setting. Let again $\phi$ be the inaccessible conceptual variable and let 
$\lambda^a=\lambda^a (\phi)$ for $a\in \mathcal{A}$ be the maximal accessible conceptual variables satisfying Assumption 1. Let the corresponding 
induced groups $G^a$ and $G$ satisfy Assumption 2. Let us assume either the spin/ angular momentum case or that Assumption 3 holds. Finally, let $t^{a}$ for each $a$ be an arbitrary real function on the range of $\lambda^a$, 
and assume that we instead of focusing on $\lambda^a$, focus on $\theta^a =t^{a}(\lambda^a)$ for each $a\in \mathcal{A}$. I will call this the 
symmetrical epistemic setting; the e-variables $\theta^a$ are no longer maximal.

Consider first how to define the quantum states $|a;k\rangle$. We are no longer interested in the full information on $\lambda^a$, but keep first the Hilbert 
space as in Section 4. For definiteness look at the case where Assumption 3 holds. Then let $f_{k}^a (\phi)=I(t^{a}(\lambda^a)=t^{a}(u_{k}))=I(\theta^a =u_{k}^a)$, where $u_{k}^a =t^{a}(u_{k})$.  We let 
again $g_{0a1}'$ and $g_{0a2}'$ be two distinct elements of $G'$ such that $g_{0ai}'\rightarrow g_{0a}$, define $g_a'=(g_{0a1}')^{-1}g_{0a2}'$ and let $g_a'\rightarrow g_a^0$ under the homomorphism of Proposition 1c). Then define
\[|a;k\rangle =V((g_a')^{-1})U(g_{0a}) f^a_k(\phi) =U((g_a^0)^{-1})f_{ka}^0(\phi)=f_{ka}^0(g_a^0\phi)\]
with $f_{ka}^0(\phi)=I(t^a(\lambda^0(\phi))=u_k^a)$.
\bigskip

\textbf{Interpretation of the state vector $|a;k\rangle$:} \textit{1) The question: 'What is the value of $\theta^a$?' has been posed. 2) We have obtained 
the answer $\theta^a =u_k^a$. Both the question and the answer are contained in the state vector.}
\bigskip

From this we may define the operator connected to the e-variable $\theta^a$:
\[A^a =\sum_k u_k^a |a;k\rangle\langle a;k|=\sum_{k} t^{a}(u_{k})|a;k\rangle\langle a;k|.\]
Then $A^a$ is no longer necessarily an operator with distinct eigenvalues, but  $A^a$ is still Hermitian: $A^{a\dagger}=A^a$.
\bigskip

\textbf{Interpretation of the operator $A^a$:} \textit{ This gives all possible states and all possible values corresponding to the accessible e-variable 
$\theta^a$.}
\bigskip

The projectors $|a;k\rangle\langle a;k|$  and hence the ket vectors $|a;k\rangle$ are no longer uniquely determined by $A^a$: They can be transformed 
arbitrarily by unitary transformations in each space corresponding to one eigenvalue. As long as the focus is only on $\theta^a$, or $A^a$, I will redefine $|a;k\rangle$ by allowing it to be subject to 
such transformations. These transformed eigenvectors all still correspond to the same eigenvalue, that is, the same observed value of $\theta^a$ and they give 
the same operators $A^a$. In particular, in the maximal symmetric epistemic setting I will allow an arbitrary constant phase factor in the functions defining the 
$|a;k\rangle$'s. 

As an important example of the general construction, assume that $\lambda^a$ is a vector: $\lambda^a =(\theta^{a_1},...,\theta^{a_m})$. Then one can 
write a state vector corresponding to $\lambda^a$ as
\[|a;k\rangle =|a_1;k_1\rangle \otimes ...\otimes |a_m;k_m\rangle\]
in an obvious notation, where $a=(a_1,...,a_m)$ and $k=(k_1,...,k_m)$. The different $\theta$'s may be connected to different subsystems. This construction can also be made, and is of great interest, in the spin/ angular momentum case. 

So far I have kept the same groups $G^a$ and $G$ when going from $\lambda^a$ to $\theta^a =t^{a}(\lambda^a)$, that is from the maximal symmetrical 
epistemic setting to the general symmetrical epistemic setting. This implies that the (large) Hilbert space will be the same. A special case occurs if $t^{a}$ is a 
reduction to an orbit of $G^a$. Then the construction of this section can also be carried 
with a smaller group action acting just upon an orbit, resulting then in a smaller Hilbert space. In the example of the previous paragraph it may be relevant to 
consider one Hilbert space for each subsystem. The large Hilbert space is however the correct space to use when the whole system is considered. 

These considerations are highly relevant when considering several observers. One single observer may have access to just a few subsystems. In addition he has 
his own context. From this context one can define what is his accessible and inaccessible conceptual variables. In the same way a group of several observers may through 
verbal communication arrive at a common context, and from this context one can define their accessible and inaccessible  conceptual variables. Imagine that these observers 
together observe a particular physical system, and consider the corresponding Hilbert space.
\bigskip

\textbf{Assumption 4.}
 \textit{For most physical system at some particular time one can either imagine an observer or a group of communicating observers 
for which the assumptions of the symmetrical epistemic setting are satisfied. In many cases all real and imagined observers agree on the physical observations, 
in which case this is considered part of the ontic world.}
\bigskip

It is important that for every physical system we can consider imagined observers.  All real 
and  imagined observers may agree on observations like charge or non-relativistic mass of a particle. For these observations we do not need the formalism of 
quantum mechanics. For other variables the construction of this section  applies.

At any time we can also imagine non-communicating observers. For each observer then the general symmetrical setting may apply. Particular state vectors in 
each observer's Hilbert space might then be linear combinations of primitive state vectors of the form $|a_{1};k_{1}\rangle \otimes ... \otimes |a_{s};k_{s}\rangle$. 
As is well known, when linear combinations of primitive state vectors can not be reduced to a primitive form, they are called entangled state vectors. Entangled state vectors play an important role in many 
discussions of quantum mechanics.

When the physical system available to an observer $A$ is subject to a non-maximal symmetrical setting resulting in an operator 
$A^a =\sum_{k}u_{k}^a |a;k\rangle \langle a;k|$, we can always imagine another observer $B$, communicating with $A$, with an operator 
$B^b =\sum_{j}v_{j}^b |b;j\rangle \langle b;j|$, so that the operator 
$A^a \otimes B^a =\sum_{k,j}u_{k}^a v_{j}^b (|a;k\rangle \otimes |b;j\rangle)(\langle a;k|\otimes \langle b;j|)$ corresponds to a maximal setting, 
i.e., has distinct eigenvalues.

Assumption 4 is assumed to hold for nearly all physical systems, and also for combinations of physical systems. Through the imagined observers the 
construction of Section 5 can be carried out, and for each case a Hilbert space can be constructed.

Connected to any general physical system, one may have several e-variables $\theta$ and corresponding operators $A$. In the ordinary quantum formalism there is well-known theorem saying that, in my formulation, $\theta^{1},...,\theta^{n}$ are compatible, that is, there exists an 
e-variable $\lambda$ such that $\theta^{i}=t^{i}(\lambda)$ for some functions $t^{i}$ if and only if the corresponding operators commute:
\[[A^{i},A^{j}]\equiv A^{i}A^{j}-A^{j}A^{i}=0\ \mathrm{for\ all}\ i,j.\]
(See Holevo [13].) Compatible e-variables may in principle be estimated simultaneously with arbitrary accuracy.

\section{Further developments}

In [9] I discuss a number of further implications of a possible Hilbert space formulation of epistemic processes.

First, the density operators are defined by
\[\sigma^a=\pi_k^a \sum_k |a;k\rangle\langle a;k|.\]
Here the $\pi_k^a$ are either prior probabilities, posterior probabilities, or derived from confidence distributions. Traditionally, statisticians are divided into Bayesians and frequentists; I want to include both cultures. Prior  distributions and posterion distributions are Bayesian concepts for parameters. Confidence distributions is a recent corresponding frequentist concept for distribution of parameters [14]. 

For simple epistemic states, Born's formula can be written as
\[ P(\theta^b=u_j^b|\theta^a=u_k^a)=|\langle a;k|b;j\rangle|^2.\]
This formula is proved in [9] from 1) An extension of the likelihood principle from statistics; 2) An assumption of rationality (Dutch book principle); 3) Paul Busch's version of Gleason's theorem [15]. 

In statistics, the likelihood is the joint density of the observation, and the likelihood principle states that observations with the same likelihood produce the same experimental evidence. This principle can be proved from fairly obvious presuppositions. The Dutch book principle states: For a rational actor, no choice of payoffs in a series of bets shall lead to sure loss for the bettor. Thus in this approach, Born's formula, which is formulated as an independent axiom in most texbooks, can be seen to follow from quite intuitive assumptions.

In [9] also the quantum mechanics for position and momentum is discussed. The Schr\"{o}dinger equation $i\hbar \frac{d}{dt}|\psi\rangle_t =H|\psi\rangle_t$ for one-dimensional position is shown to follow from 1) Position as an inaccessible stochastic process; 2) Conditioning both on past events and on future events; 3) Elements from Nelson's stochastic mechanics.

EPR and violation of Bell's inequality for simple epistemic states can be given interesting interpretations. The conditionality principle from statistics states: When doing a coin toss to choose an experiment, an observer should condition upon the value of the coin toss. Imagine that this principle is used both for Alice and for Bob. This implies that their observations are interpreted as some sort of experiment. That is, observation means inference: Decision on what has been observed. Under this condition the EPR paradox turns out to be no paradox any more, and the violation of Bell's inequality can be simply explained. The assumption of \emph{realism} is violated by this version of quantum mechanics; the assumption of locality is not violated.

It is well known from the Quantum Bayesian literature that the so-called paradox of Wigner's friend is no paradox any more with an epistemic interpretation of the state vector.

\section{Concluding remarks}

Despite its enormous success, there is still no consensus among physicists about what quantum theory is saying about the nature of reality. In my view, its foundation should be related to the process of obtaining knowledge about the world. It seems to be possible to formalize this in a precise mathematical way. 

In this theory something resembling the formal state definition of traditional quantum mechanics is derived from a simpler and more natural conceptual basis. This derivation is done under certain technical assumptions. The use of group theory seems to be essential.

It also seems to be important that the approach here and in [9] uses ideas both from the quantum mechanical tradition and from the statistical tradition. In the long run one might perhaps hope for some sort of synthesis in the foundation of the two sciences.

Since space is limited here, I prepare an extended version [16], where I will discuss the validity of Theorem 1 for the spin/angular momentum case, and where the relation to other approaches towards the foundation of quantum mechanics and to the interpretation of quantum mechanics will be discussed.

\section*{References}

\begin{description}

\item[]
[1]	Schlossbauer~M, Koller~J and Zellinger~A 2013  A snapshot of fundamental attitudes toward quantum mechanics (\textit{Preprint} quant-ph/1301.1069)

\item[]
[2]	Norsen~T and Nelson~S 2013 Yet another snapshot of fundamental attitudes toward quantum mechanics (\textit{Preprint} quant-ph/1306.4646)

\item[]
[3]     Spekkens~R~W 2007 \textit{Phys. Rev.} A \textbf{75} 032110

\item[]
[4]  Fuchs~C~A 2010 QBism, the Perimeter of Quantum Bayesianism (\textit{Preprint} quant-ph/1003.5209)

\item[]
[5]  Fuchs~C~A and Schack~R 2011 \textit{Foundations of Physics} \textbf{41}, 345

\item[]
[6] Fuchs~C~A, Mermin~N~D and Schack~R 2013 An introduction to QBism with application to the locality of quantum mechanics (\textit{Prepront} quant-ph/1311.5253)

\item[]
[7] Timpson~C~G 2008 \textit{Studies in History and Philosophy of Modern Physics} \textbf{39} 579

\item[]
[8] von Baeyer~H~C 2013 \textit{Scientific American} \textbf{308} (6), 38

\item[]
[9] Helland~I~S In preparation \textit{Epistemic Processes. A Basis for Statistics and for Quantum Mechanics.} Book manuscript

\item[]
[10] Fisher~ R~A 1922 On the mathematical foundation of theoretical statistics. Reprinted in Fisher~R~A 1950 \textit{Contribution to Mathematical Statistics} (New York: Wiley)

\item[]
[11] Helland~I~S 2004 \textit{International Statistical Review} \textbf{72}, 409

\item[]
[12] Helland~I~S  2010 \textit{Steps Towards a Unified Basis for Scientific Models and Methods} (Singapore: World Scientific)

\item[]
[13] Holevo~A~S 1982 \textit{Probabilistic and Statistical Aspects of Quantum Theory} (Amsterdam: North Holland)

\item[]
[14] Schweder~T and Hjort~N~L 2002 \textit{Scandinavian Journal of Statistics} \textbf{29}, 309

\item[]
[15] Busch~P 2003 \textit{Physical Review Letters} \textbf{91} (12), 120403

\item[]
[16] Helland~I~S 2015 Approach towards the quantum formulation from assumptions of epistemic processes. Article in preparation

\end{description}

\end{document}